\documentclass[twocolumn,showpacs,preprintnumbers,amsmath,amssymb,superscriptaddress]{revtex4}

\usepackage{graphicx}
\usepackage{graphicx}
\usepackage{here}
\begin{document}
\title{Effect of strong localization of doped holes 
in angle-resolved photoemission spectra of La$_{1-x}$Sr$_x$FeO$_3$}

\author{H. Wadati}
\affiliation{Department of Physics and Department of Complexity 
Science and Engineering, University of Tokyo, 
Kashiwa, Chiba 277-8561, Japan}

\author{A. Chikamatsu}
\affiliation{Department of Applied Chemistry, University of Tokyo, 
Bunkyo-ku, Tokyo 113-8656, Japan}

\author{M. Takizawa}
\affiliation{Department of Physics and Department of Complexity 
Science and Engineering, University of Tokyo, 
Kashiwa, Chiba 277-8561, Japan}

\author{R. Hashimoto}
\affiliation{Department of Applied Chemistry, University of Tokyo, 
Bunkyo-ku, Tokyo 113-8656, Japan}

\author{H. Kumigashira}
\affiliation{Department of Applied Chemistry, University of Tokyo, 
Bunkyo-ku, Tokyo 113-8656, Japan}

\author{T. Yoshida}
\affiliation{Department of Physics and Department of Complexity 
Science and Engineering, University of Tokyo, 
Kashiwa, Chiba 277-8561, Japan}

\author{T. Mizokawa}
\affiliation{Department of Physics and Department of Complexity 
Science and Engineering, University of Tokyo, 
Kashiwa, Chiba 277-8561, Japan}

\author{A. Fujimori}
\affiliation{Department of Physics and Department of Complexity 
Science and Engineering, University of Tokyo, 
Kashiwa, Chiba 277-8561, Japan}

\author{M. Oshima}
\affiliation{Department of Applied Chemistry, University of Tokyo, 
Bunkyo-ku, Tokyo 113-8656, Japan}

\author{M. Lippmaa}
\affiliation{Institute for Solid State Physics, University of Tokyo, 
Kashiwa, Chiba 277-8581, Japan}

\author{M. Kawasaki}
\affiliation{Institute for Materials Research, Tohoku University, 
2-1-1 Katahira, Aoba, Sendai 980-8577, Japan}

\author{H. Koinuma}
\affiliation{National Institute for Materials Science, 
1-2-1 Sengen, Tsukuba 305-0047, Japan}

\date{\today}
\begin{abstract}
We have performed an angle-resolved 
photoemission spectroscopy study of 
La$_{0.6}$Sr$_{0.4}$FeO$_3$ using {\it in situ} 
prepared thin films 
and determined its band structure. 
The experimental band dispersions could be well explained by 
an empirical band structure assuming the G-type 
antiferromagnetic state. However, 
the Fe $3d$ bands were found to be 
shifted downward relative 
to the Fermi level ($E_F$) 
by $\sim 1$ eV compared with the calculation 
and to form 
a (pseudo)gap of $\sim 1$ eV at $E_F$. 
We attribute this observation to a strong localization 
effect of doped holes due to polaron formation. 
\end{abstract}
\maketitle
Metal-insulator (MI) transitions in strongly-correlated electron 
systems have been generally understood in terms of band-width control and 
filling control \cite{rev}. However, the actual situation is more
complicated because of the effect of disorder, electron-phonon
interaction, charge and orbital ordering, etc \cite{KM,Rosch,Mn,VO2,Fe3O4}. 
Electron-phonon interaction causes a polaronic effect on 
charge carriers and increases their effective masses. 
In some filling-controlled systems, 
carriers doped into the Mott insulator 
remain localized as self-trapped small polarons 
and the system remains insulating. 
The problem of how an MI transition occurs 
when strong electron-phonon interaction is present 
remains highly controversial. 
A striking example is a hole-doped Mott insulator 
(charge-transfer-type insulator) 
La$_{1-x}$Sr$_x$FeO$_3$ (LSFO), where 
the insulating phase is unusually wide in the phase diagram 
($0<x<0.5$ at room temperature 
and $0<x<0.7$ at low temperatures) \cite{Matsuno}. 
In a previous photoemission study, the gap at the Fermi level ($E_F$) 
was seen for all compositions in the range 
$0\le x \le 0.67$ \cite{Wadati}, 
consistent with the wide insulating region. 

The effect of electron-phonon interaction is 
clearly reflected on the spectral function 
\cite{Mahan}, and recent 
photoemission results have 
been discussed in this context for 
high-$T_c$ superconductors \cite{KM,Rosch}, 
Mn oxides \cite{Mn}, 
VO$_2$ \cite{VO2}, and Fe$_3$O$_4$ \cite{Fe3O4}. 
Angle-resolved photoemission spectroscopy (ARPES)
is a particularly powerful technique 
by which one can directly study 
the band structure of a material. 
However, there have been few studies on transition metal oxides 
(TMOs) with the 
cubic perovskite structures like LSFO because many of them 
do not have a cleavage plane. 
Recently, the band structure of 
the three-dimensional perovskite
La$_{1-x}$Sr$_x$MnO$_3$ has been studied by ARPES 
using well-defined surfaces of 
{\it in situ} prepared epitaxial 
thin films \cite{Shi,Chika}, 
demonstrating that such an approach  
on TMO films is one of the best methods 
for such a purpose. 
In this paper, we report on 
{\it in-situ} ARPES results of single-crystal LSFO
($x=0.4$) thin films grown on SrTiO$_3$ $(001)$ substrates. 
We observed several dispersive bands which can be assigned 
to the Fe $3d$ $e_g$, $t_{2g}$ and O $2p$ bands. 
The experimental band structure was interpreted 
based on an empirical tight-binding band structure 
assuming the G-type antiferromagnetic (AF) state. 
In spite of the overall agreement between experiment and theory, 
fundamental discrepancies were found for the overall shift of 
the Fe $3d$ bands away from $E_F$ and 
the suppression of 
spectral weight around $E_F$. 
We shall discuss 
these discrepancies in terms of strong polaronic effect.

Experiment was carried out using a photoemission
spectroscopy (PES) system combined with a laser molecular 
beam epitaxy (MBE) 
chamber at beamline BL-1C of the
Photon Factory, KEK \cite{Horiba}. 
The LSFO $(x=0.4)$ thin films 
were grown epitaxially on Nb-doped SrTiO$_3$ 
substrates by the pulsed laser deposition method. 
Details are described in Ref.~\cite{Wadati}. 
The lattice constants were determined 
to be 
$a=b=3.905\ \mbox{\AA}$ and $c=3.883 \ \mbox{\AA}$. 
In this paper, 
$k_{\parallel}$ denotes the in-plane electron momentum, 
and $k_z$ the out-of-plane one, 
expressed in units of $\pi/a$ or $\pi/c$. 
By low energy electron diffraction, 
sharp 1~1 spots were observed with no sign 
of surface reconstruction. 
ARPES measurements were performed under 
ultrahigh vacuum of $\sim 10^{-10}$ Torr at 150 K, 
below the N\'eel temperature ($T_N=320$ K) of LSFO with 
$x=0.4$ \cite{Matsuno}, 
using a Scienta SES-100 electron energy analyzer. The 
total energy resolution was set to about 150 meV. 
The $E_F$ position was determined by measuring gold
spectra. 

\begin{figure}
\begin{center}
\includegraphics[width=7.5cm]{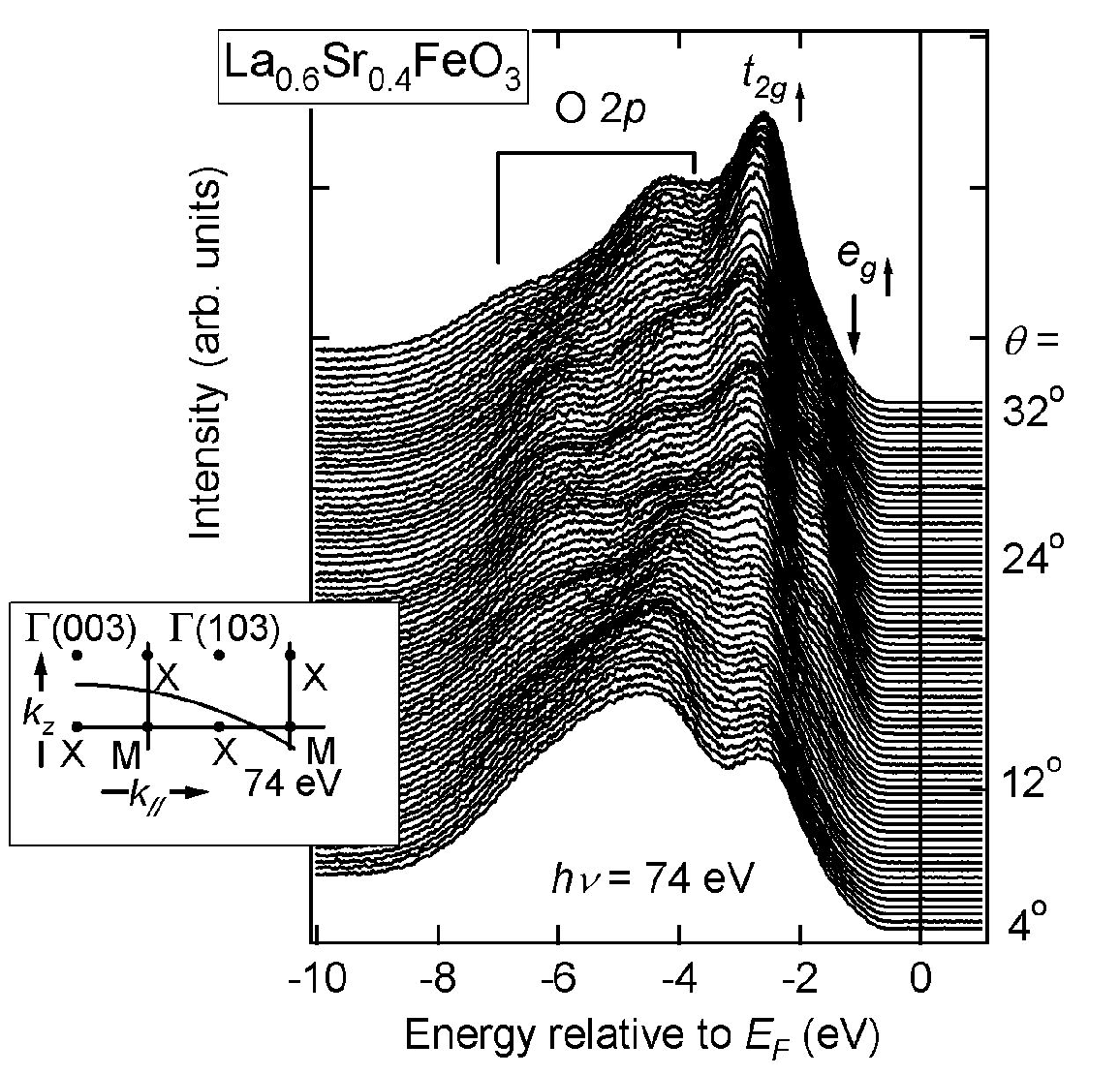}
\caption{ARPES spectra of an LSFO ($x=0.4$) thin film
taken at $h\nu =74$ eV. Polar angle ($\theta$) referenced 
to the surface normal is indicated. The inset shows the trace in $k$-space.}
\label{ARPES}
\end{center}
\end{figure}

Figure \ref{ARPES} shows ARPES spectra of an LSFO ($x=0.4$) 
thin film taken 
with a photon energy of 74 eV. 
The inset shows a trace in $k$-space 
for $h\nu =$ 74 eV with 
changing polar emission angle $\theta$. 
Here, we have assumed the work function of the 
sample $\phi = 4.5$ eV, 
and the inner potential $V_0=10.5$ eV so that 
the periodicity of the band dispersion is correctly given 
in normal emission (not shown). 
The shoulder structure at $\sim -1.3$ eV 
shows a significant dispersion, and 
is assigned to the Fe $3d$ majority-spin $e_g$ bands. 
The peak structure at $\sim -2.4$ eV shows a weaker dispersion, 
and is assigned to 
the Fe $3d$ $t_{2g}$ majority-spin bands. 
The structures at $-(4 -7)$ eV show strong angular dependence 
and are assigned to O $2p$ bands. 
The $e_g$ bands do not cross $E_F$, consistent with the insulating 
behavior and the persistence of the gap observed in the 
angle-integrated PES (AIPES) 
spectra \cite{Wadati}. 

\begin{figure}
\begin{center}
\includegraphics[width=9cm]{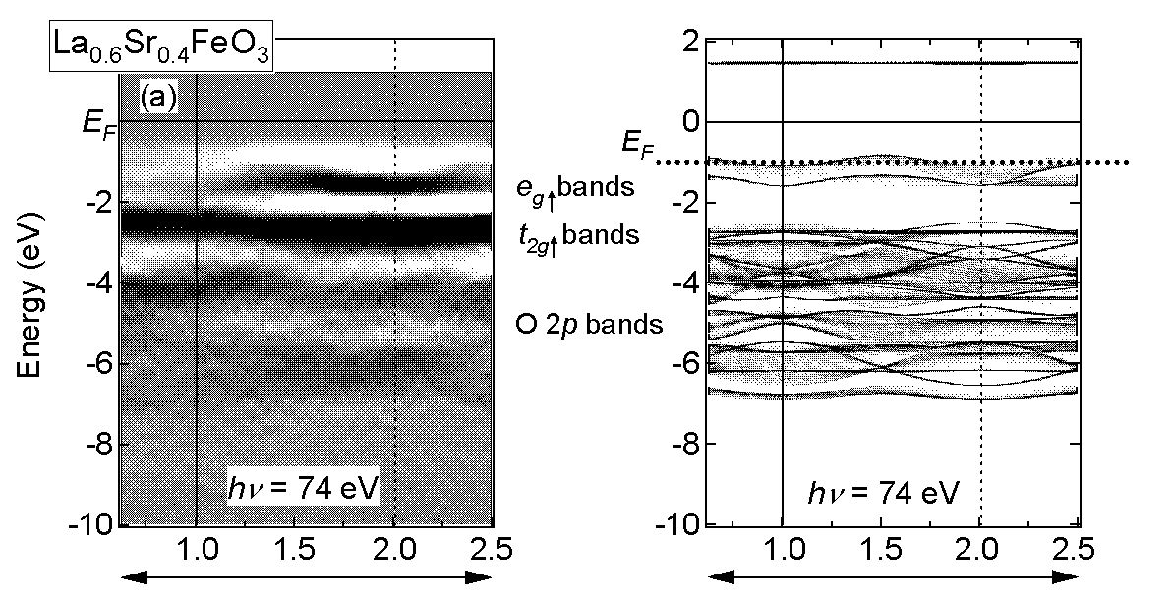}
\includegraphics[width=9cm]{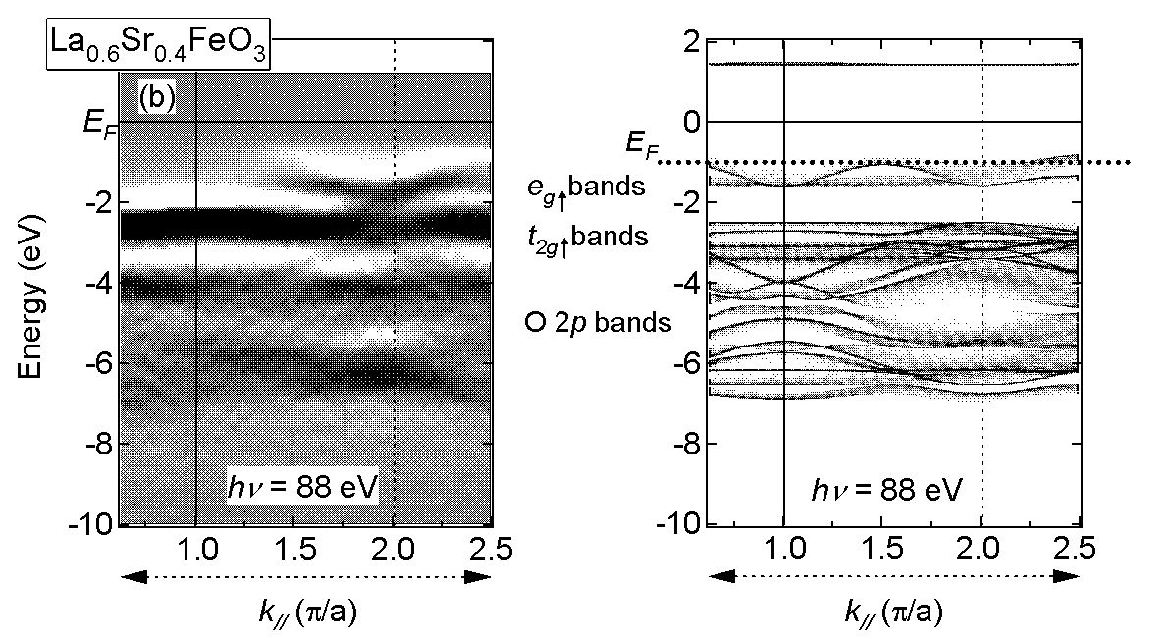}
\includegraphics[width=7cm]{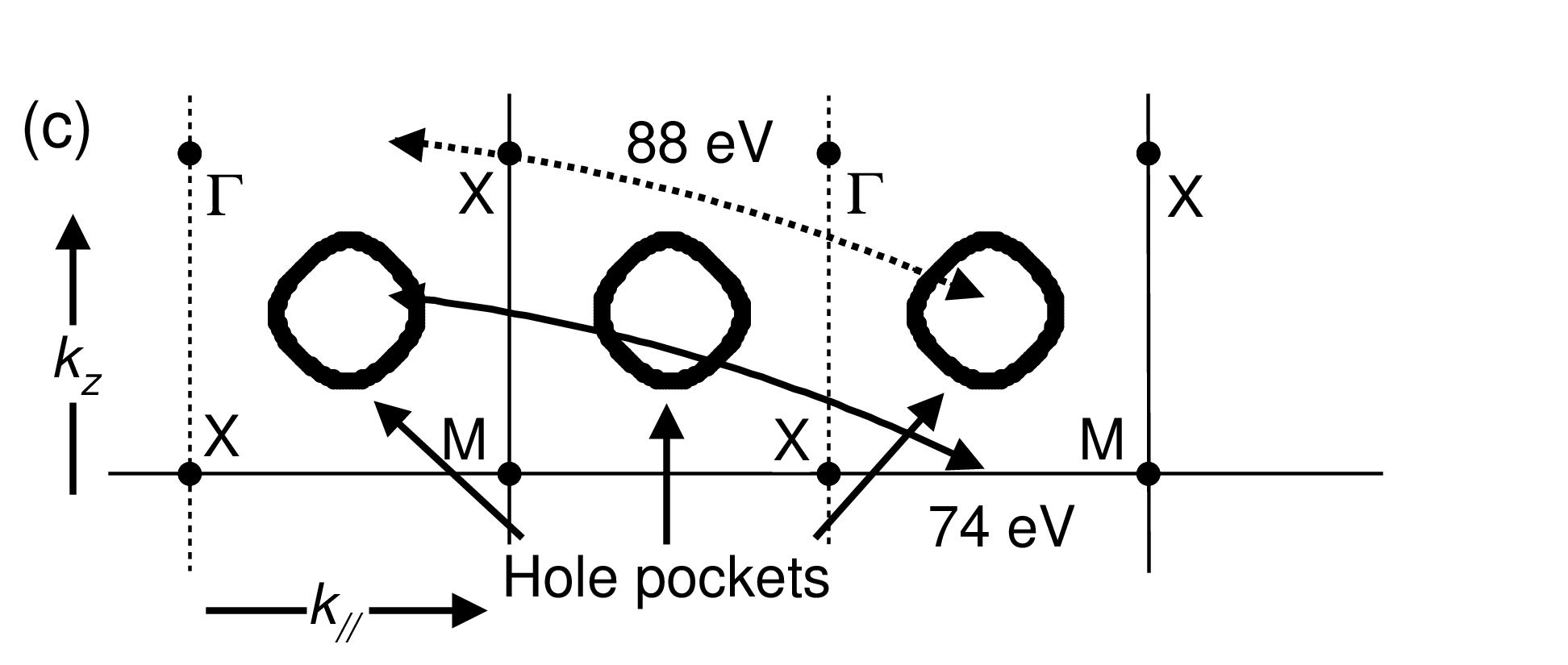}
\caption{Comparison of the ARPES spectra of LSFO 
taken at 74 eV (a) and 88 eV (b) with tight-binding calculation. 
The left panels show the experimental band structure 
deduced from the second derivatives of the EDCs 
(dark parts correspond to energy bands) and 
the right panels show the result of tight-binding calculation, 
taking into account the finite photoelectron mean-free path. 
The traces in $k$-space and calculated hole pockets are shown in (c).}
\label{tight2}
\end{center}
\end{figure}

\begin{figure}
\begin{center}
\includegraphics[width=9cm]{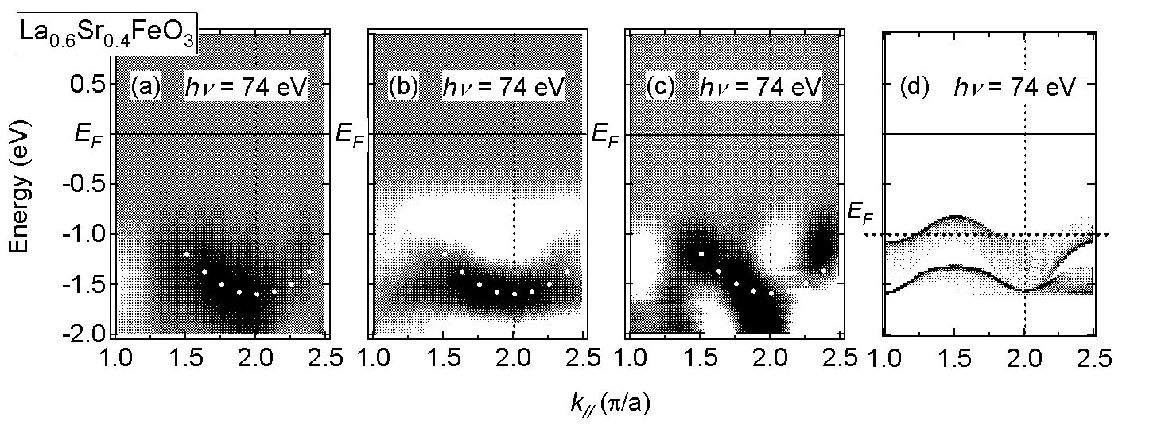}
\caption{Enlarged plot of ARPES spectra near $E_F$ in $E$-$k$ space. 
Panel (a): Intensity plot of the EDCs. 
Angle-independent part has been subtracted as a background. 
(b): Second derivatives of the EDCs. 
(c): Second derivatives of the MDCs.  
(d): Result of tight-binding calculation. 
White dotted curves are the guide to the eye.}
\label{MDC}
\end{center}
\end{figure}

The band dispersions are more clearly seen in the 
gray-scale plot 
in the left panel of Fig.~\ref{tight2} (a). 
Here, the second derivatives of the energy distribution 
curves (EDCs) are plotted on the gray scale with 
dark parts corresponding to energy bands. 
The same plot for $h\nu = 88$ eV [corresponding to 
the upper trace in Fig.~\ref{tight2} (c)] is shown in the left panel of 
Fig.~\ref{tight2} (b). 
In Fig.~\ref{tight2} (c), 
hole pockets obtained by the tight-binding calculation (described below) 
are also shown. The trace for 74 eV crosses the 
calculated hole pocket at $k_{\parallel}\sim 1.5\pi /a$, 
while that for 88 eV does not in the 
same $k_{\parallel}$ region. 
Figure \ref{MDC} shows enlarged gray-scale plots near $E_F$ 
in $E$-$k$ space. Panel (a) is 
a direct intensity plot of the EDCs in Fig.~\ref{ARPES}. 
Angle-independent part has been subtracted as a background. 
Second derivatives of the EDCs are produced in panel (b) 
and those of momentum distribution curves (MDCs) 
in panel (c). In all the plots, the band has a minimum at 
$k_{\parallel}\sim 2.0\pi /a$, and disperses 
upward toward both sides 
and disappears at $k_{\parallel}\sim 1.5\pi /a$ and 
$k_{\parallel}\sim 2.5\pi /a$, 
not at $E_F$ but $\sim 1$ eV below it.

In order to interpret the experimental band dispersions 
more quantitatively, 
we performed a tight-binding band-structure calculation 
with empirical parameters, following the scheme in 
Refs.~\cite{Kahn,ReO3,STO}. 
We performed the calculation, however, 
assuming the G-type AF state 
corresponding to the magnetic ordering in LSFO. 
Here, the effect of G-type antiferromagnetism was taken 
into account 
phenomenologically by imposing an energy difference 
$\Delta E$ between the spin-up and spin-down Fe sites. 
Parameters to be fitted were therefore $\Delta E$, 
the energy difference between the Fe $3d$ level and the 
O $2p$ level, $\epsilon_p-\epsilon_d$, 
and a Slater-Koster parameter, $(pd\sigma)$ \cite{JC}. 
For the other Slater-Koster parameters, we assumed 
$(pd\sigma)/(pd\pi)=-2.2$ \cite{WA,STO1}, 
and $(pp\sigma)$ and $(pp\pi)$ were taken to be 
0.60 eV and $-0.15$ eV, respectively \cite{STO1}. 
$(pd\sigma)$ is expected in the range of $-(1.4-1.9)$ eV 
from a configuration-interaction (CI) cluster-model calculation
\cite{Bocquet,Wadati}. 
Crystal-field splitting $10Dq$ of 0.41 eV 
was taken from Ref.~\cite{Wadati}. 
It should be noted that $\epsilon_p-\epsilon_d$, 
$(pd\sigma)$ and $\Delta E$ primarily determine 
the Fe $3d$ $-$ O $2p$ band positions, their dispersions, 
and the optical band gap, respectively, and therefore 
can be rather uniquely determined. 

The best fit to the observed band dispersions of LSFO 
and the optical gap 2.1 eV of LaFeO$_3$ (LFO) \cite{opt1} 
as shown in the right panels of 
Fig.~\ref{tight2} (a) and (b) has been obtained for 
reasoanble parameter values 
$\epsilon_p-\epsilon_d=0$ eV, 
$(pd\sigma)=-1.5$ eV, and 
$\Delta E=5.3$ eV. For these plots, 
we have considered the effect of $k_z$ broadening ($\Delta k_z$) 
caused by a finite escape depth $\lambda$ of 
photoelectrons \cite{foot} 
[$\Delta k_z \sim 1/\lambda$ ($\sim$ 0.2 $\mbox\AA^{-1}$) 
is approximately 10 \% of the Brillouin zone ($2\pi /c$)]. 
The dispersions of the $e_g$ bands 
were thus successfully reproduced. 
The weak dispersions of the $t_{2g}$ bands and the width of 
the O $2p$ bands were also well reproduced by these parameters. 
According to the band-structure calculation, spectral weight should be
cut off above the calculated $E_F$, which is determined 
by the band filling for $x=0.4$, 
while in experiment it gradually decreases 
from the calculated $E_F$ toward the experimental $E_F$. 
This discrepancy inevitably arises from the fact 
that this material is insulating up to 70 
\% hole doping while the rigid-band model 
based on the present band strucure gives the metallic state. 

\begin{figure}
\begin{center}
\includegraphics[width=9cm]{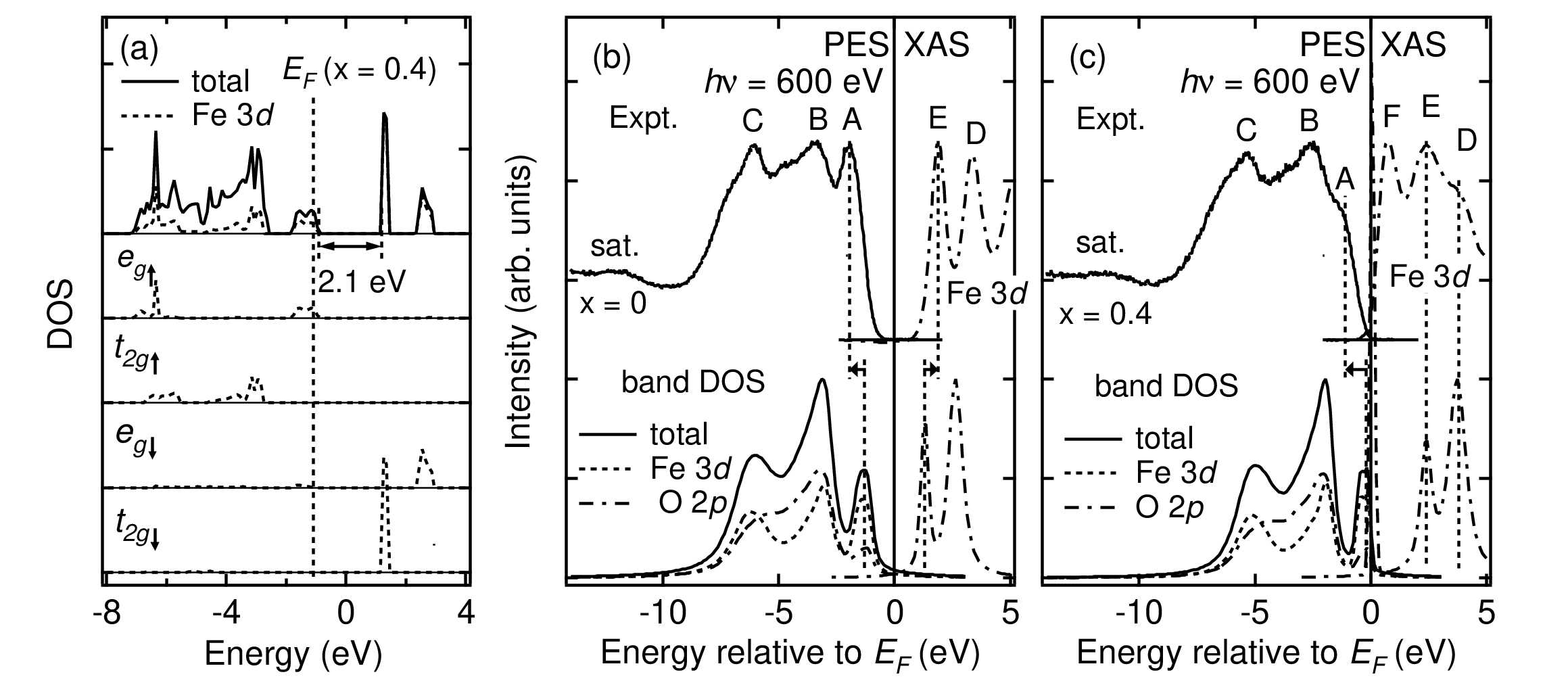}
\caption{Density of states obtained  
by using tight-binding calculation (a) and 
its comparison with the PES and O $1s$ XAS 
spectra of LFO (b) 
and LSFO ($x=0.4$) (c) taken from Ref.~\cite{Wadati}.}
\label{DOS}
\end{center}
\end{figure}

Figure \ref{DOS} (a) shows 
the density of states (DOS) of the G-type AF state 
calculated using the same parameter set. 
The partial DOS's for 
the majority-spin Fe $e_g$, 
the majority-spin $t_{2g}$, 
the minority-spin $e_g$, 
and the minority-spin $t_{2g}$ orbitals are shown 
in the lower panels. 
A large band gap opens between the occupied majority-spin $e_g$ 
bands and the unoccupied minority-spin $t_{2g}$ bands 
for the present $\Delta E$ value. 
These characteristic features were already 
reported by the previous 
Hartree-Fock calculation \cite{HF} and the local 
spin-density-approximation 
calculation \cite{Sarma1}. 
In Figs.~\ref{tight2} and \ref{DOS}, 
$E_F$ for $x=0.4$ has been determined from 
the filling of electrons when holes are doped into the AF insulator. 
In Figs.~\ref{DOS} (b) and (c) the calculated results are compared 
with the combined AIPES and O $1s$ x-ray absorption 
(XAS) spectra of LFO and 
LSFO ($x=0.4$), respectively \cite{Wadati}. 
For LFO, 
the three main structures of the valance band, 
A, B, and C are successfully 
reproduced by the calculation, while 
the satellite structure in the valence-band photoemission
spectrum could not be reproduced. 
The crystal-field splitting into 
D and E 
above $E_F$ is also well reproduced. 
For LFO, although the calculated band gap has been adjusted 
to the experimental value, the peak 
positions of structures A and E 
are still shifted away from $E_F$ in the experiment compared to the calculation 
[as denoted by broken lines in Fig.~\ref{DOS} (b)]. 
As for $x=0.4$ [Fig.~\ref{DOS} (c)], 
strucure A is shifted away from $E_F$ 
compared to the calculation even more. 

Generally speaking, doped holes may be localized due to 
disorder or through coupling to lattice distortion. 
The latter effect, namely, 
polaronic effect was recently observed in 
the photoemission spectra of a number of 
TMOs \cite{KM,Mn,VO2,Fe3O4}. 
Note that for $x=0.4$ a  hole-induced peak F appears 
within the band gap of LFO and accomodates doped holes. 
These observations cannot be explained by the 
rigid-band model in which $E_F$ has been shifted 
according to the band filling. 
The breakdown of the rigid-band model means that 
doped holes do not enter the top of the $e_g$ majority-spin band 
but enter localized states split off from the top of the $e_g$ band, 
causing the insulating behavior of this material. 

\begin{figure}[H]
\begin{center}
\includegraphics[width=8cm]{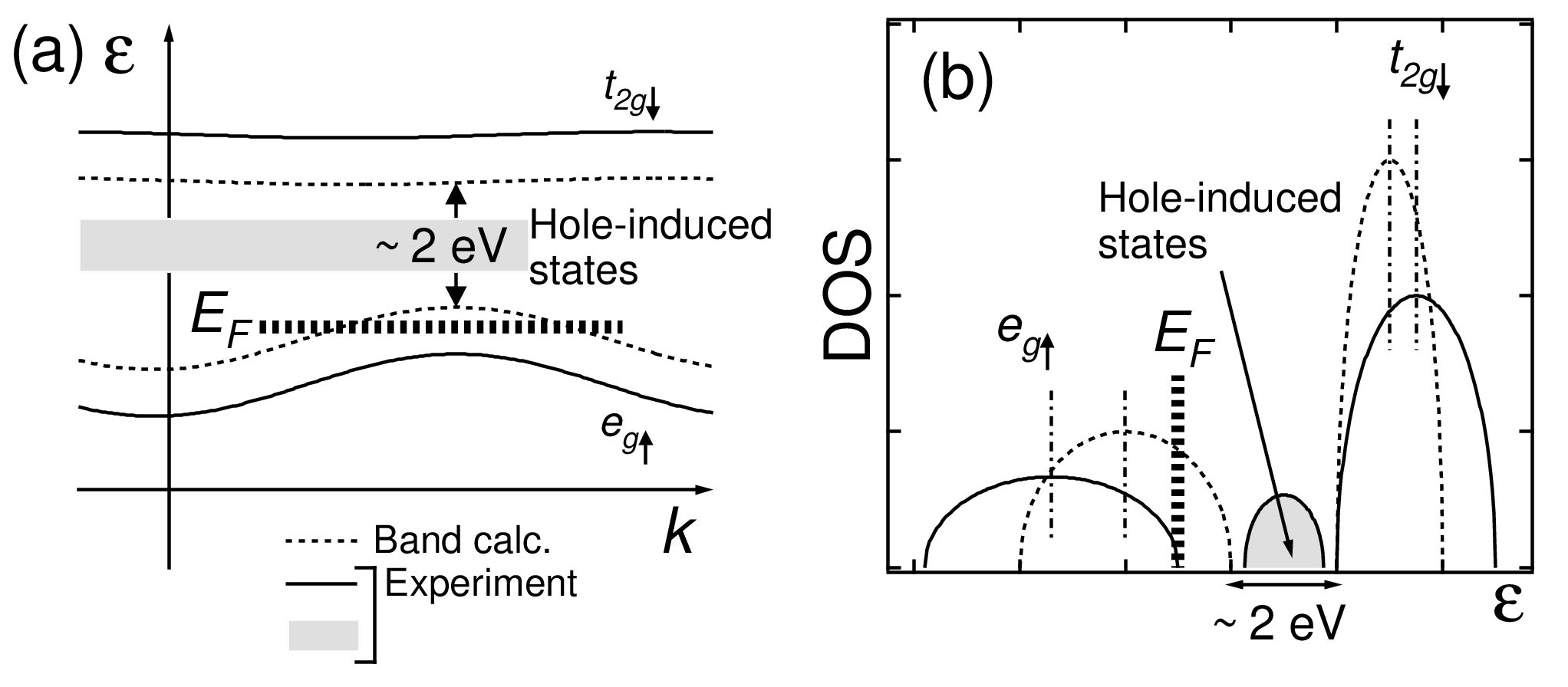}
\caption{Schematic picture of polaronic effects and hole-induced 
states in LSFO. (a) Band dispersions, (b) DOS.}
\label{tight3}
\end{center}
\end{figure}

Finally, we further 
discuss the discrepancy between 
the band-structure calculation and the experimental results 
near $E_F$ 
and the origin of the insulating behavior of LSFO. 
Figure \ref{tight3} schematically shows the calculated and experimental 
band structure and DOS of LSFO. 
Our tight-binding calculation is similar to local-density approximation 
(LDA) $+U$ in the sense that the effect of electron-electron interaction 
and hence the value of the band gap is adjusted via $\Delta E$ in the
tight-binding calculation and via $U$ in the LDA $+U$ calculation. 
Broken lines indicate the tight-binding or LDA $+U$ band 
structure, where the value of the band gap of the parent insulator 
has been adjusted to $\sim 2$ eV in order to reproduce the optical gap
\cite{opt1}. 
When holes are doped into this system, $E_F$ moves 
downward and crosses the $e_{g\uparrow}$ band, making the system metallic if 
the rigid-band model can be applied. 
However, from the comparison of 
calculation and experiment in Figs.~\ref{tight2} and \ref{DOS}, 
the $e_{g\uparrow}$ structures are shifted away from $E_F$, 
and hole-induced states appear above $E_F$. 
The existence of these hole-induced states means that 
the rigid-band model is no more valid, and 
doped holes 
enter split-off localized states formed by 
hole doping. 
There is also 
a spectral line-shape broadening compared to the 
band-structure calculation. 
Such a modification of the structures and line-shape broadening have been 
attributed to 
a polaronic effect recently observed in a number of 
TMOs \cite{KM,VO2,Fe3O4}. For example, 
in the case of Ca$_{2-x}$Na$_x$CuO$_2$Cl$_2$, the spectral 
weight of quasiparticle peak $Z$ becomes extremely small 
due to a strong coupling to phonons \cite{KM}. In the 
case of VO$_2$ \cite{VO2} and 
Fe$_3$O$_4$ \cite{Fe3O4}, spectral changes with temperature 
were interpreted by considering strong coupling 
of electrons to phonons. In the case of LSFO, 
we consider that electron-phonon coupling 
is strong enough to 
make $Z$ almost zero (an insulating behavior), and shift 
the spectral features 
away from $E_F$. Since electron-phonon coupling 
is very strong for the $e_g$ orbitals compared 
to the $t_{2g}$ orbitals, the effect can be dramatic 
as observed in LSFO, where doped holes have $e_g$ 
character. 
Moreover, according to the temperature-dependent 
PES and XAS studies, short-range (local) 
charge order may exist in the wide composition range \cite{Wadati2}. 
The polaronic effect probably enhances the tendency toward local charge 
ordering and explains the unusually wide insulating phase in LSFO. 

In summary, we have performed an {\it in-situ} ARPES 
study on single-crystal 
LSFO ($x=0.4$) thin films 
and determined its band structure. 
The observed dispersive bands were assigned to 
Fe $3d$ $e_g$, $t_{2g}$ and O $2p$ bands. 
The experimental band dispersions were interpreted 
using a tight-binding band-structure calculation 
by assuming 
the G-type antiferromagnetic state. 
From the discrepancy near $E_F$ between experiment and 
theory in spite of the overall agreement, 
the insulating behavior of this material is 
proposed to be caused by the localization of hole-induced 
states due to polaronic effect and/or 
short-range charge order. 

The authors would like to thank 
N. Hamada and T. Saha Dasgupta for informative discussion and 
D. Kobayashi for sample preparation. 
We are also grateful to K. Ono for their support at KEK-PF. 
This work was
supported by a Grant-in-Aid for Scientific Research
(A16204024) from the Japan Society for the Promotion of Science (JSPS). 
This work was done under the approval of the Photon Factory 
Program Advisory Committee (Proposal No. 2002S2-002). 
H. W. acknowledges financial support from JSPS. 

\begin{thebibliography}{10}
\bibitem{foot}
The effect of $k_z$ broadening is given by the following 
Lorentzian function, 
$L(k_z)=1/\{(k_z-k_{z0})^2+(1/2\lambda)^2\}$, 
where $\lambda$ denotes an escape depth of photoelectrons. 
Here we used the value $\lambda=5$ $\mbox\AA$.
\end{thebibliography}

\end{document}